# Urban-Semantic Computer Vision:
# A Framework for Contextual Understanding of People in Urban Spaces

**Anthony Vanky, Ph.D., Corresponding author**
Graduate School of Architecture, Planning and Preservation
Columbia University, New York, NY 10025
Email: a.p.vanky@columbia.edu

**Ri Le**
Graduate School of Architecture, Planning and Preservation
Columbia University, New York, NY 10025
Email: r.le@columbia.edu

**ABSTRACT**

Increasing computational power and improving deep learning methods have made computer vision technologies pervasively common in urban environments. Their applications in policing, traffic management, and documenting public spaces are increasingly common. Despite the often-discussed biases in the algorithms' training and unequally borne benefits, almost all applications similarly reduce urban experiences to simplistic, reductive, and mechanistic measures. There is a lack of context, depth, and specificity in these practices that enables semantic knowledge or analysis within urban contexts, especially within the context of using and occupying urban space. This paper will critique existing uses of artificial intelligence and computer vision in urban practices to propose a new framework for understanding people, action, and public space.

This paper revisits Geertz's use of thick descriptions in generating interpretive theories of culture and activity and uses this lens to establish a framework to evaluate the varied uses of computer vision technologies that weigh meaning. We discuss how the framework's positioning may differ (and conflict) between different users of the technology. This paper also discusses the current use and training of deep learning algorithms and how this process limits semantic learning and proposes three potential methodologies for gaining a more contextually specific, urban-semantic, description of urban space relevant to urbanists.

This paper contributes to the critical conversations regarding the proliferation of artificial intelligence by challenging the current applications of these technologies in the urban environment by highlighting their failures within this context while also proposing an evolution of these algorithms that may ultimately make them sensitive and useful within this spatial and cultural milieu.

# 1. INTRODUCTION

Ever-powerful computational ability, the reduced cost of communications infrastructure, and the increase-diminishing size of sensors have enabled the pervasive placement of technologies into the fabric of urban spaces, birthing a movement of the "smart city." For many in the so-called smart cities movement, the trend has been towards the instrumentalization of cities, finding greater efficiencies, and the problematization of many facets of urban living (Hollands, 2008). For technologists working in this field, the enumeration game is being applied to all domains ranging from mobility and infrastructure to public safety and democratic participation (Eagle & Pentland, 2009; Goldsmith & Crawford, 2014; Jiang et al., 2013). Nevertheless, the defining social characteristics of urban space defy a reduction to a mere optimization problem.

These digital technologies and their resultant models and data outcomes have the ability to shape our perspective of the built environment. No different than the well-publicized challenges of bias prevalently found in other algorithmic processes (Crawford, 2018; Kirchner et al., 2016), the black box methodologies and opaque outcomes so too can unduly influence our reading of the places we inhabit (Schwarzer, 2017). Despite this conflict between reductivism and complexity, there is an urgent need to understand through new models and tools may open new avenues for research into public space and urban form in light of rapid urbanization and the increased privatization of urban space (E. Talen & Ellis, 2002).

As if it were an iteration from the modernist use of photography in urban planning, the growing ubiquity of deep learning and computer vision applications have created new opportunities to understand cities through imagery (Lecun et al., 2015). While there is optimism in how these emerging technologies can allow for a more precise (and perhaps, broad) method for understanding cities, questions remain.

Within this computational milieu, this paper focuses specifically on the nascent, but growing role of these algorithmic tools are being applied to urban planning and management. In one sense, they quantify human behavior in urban space that offers the ability for decision-makers to base policy in more informed ways. In another, this numerical reductivism applied to urban space is blind to the specificity and essential character that makes cities unique places of inhabitation. As such, this paper argues that in addition to situated technologies' reductivist orientation, there is a need for a distinct approach to their use in the understanding of how people inhabit and use these public urban spaces. Further, with the increased proliferation of computer-vision and image-based approaches toward the instrumentalization of cities, an urban-specific lexicon to the training, implementation, and adoption of urban technologies.

While the use of computer vision technologies has spanned many facets of urban management, such as infrastructure utilization and public safety, this paper considers explicitly using these technologies to understand how people inhabit and use public space. This paper argues that image-based artificial intelligence is a natural progression in the modernist use of photo imagery to capture data on the city. This paper also reviews, in brief, how artificial intelligence is being applied to image-based data to illustrate the potential weaknesses in creating thicker, domain-specific ontologies about the occupants and the spaces in which they inhabit. Further, this paper discusses how thin data is being marketed by both public and private actors to the potential detriment of planning human-centric spaces. However, this paper also discusses potential approaches to reconceptualize the methodologies currently used to get toward an urban-semantic description of public space using these algorithmically-based methodologies despite these limitations.

# 2. "URBANISM", AND THE ISSUE OF CONTEXT

A fundamental challenge is defining what characteristics make a space or practice *urban*, especially when contemporary city building is neo-liberal and capital driven. In other words, what makes Washington Square Park *urban* while the Hudson Yards feel not, despite being just a few kilometers apart in Manhattan? Or what makes Seoul's Namdaemun Night Markets an *urban* experience versus the nearby, impressively large Lotte Shopping Center?

By their nature, urban space is where individual experience comes together with strangers, even though they seldom share our values, history, and perspective. It is in these spaces where the mingling and contact with individuals and groups differ in their social presentation, appearance, and experience. As urbanist Michael Brill (1989) frames these experiences, it is where inhabitants "can seek and find excitement and extraordinariness in the productions and presentations created by strangers, and in those they create themselves." The spaces around them influence these social dynamics; the form and configuration of urban space frame the interactions of those in them. Ultimately, public life is uniquely able to be experienced in these spatial commons—the streets and public, urban spaces—because of how they make inhabitants' acute awareness of their dependence on one another, as well as the societal obligations that dependence spawned (Chidster, 1989). Cities are, in a sense, where individualism intermixes to create collective experience.

Defining these socio-spatial interrelationships is a difficult challenge, and there exist differing perspectives on the nature of a good public realm. There exist divergent but parallel theories about the character of these spaces, which take up a significant portion of the physical city: streets, sidewalks, squares, arcades, non-motorized transport, greenways, and the like. Jane Jacobs (1961) describes the importance of streets as public spaces:

> *"Streets are almost always public: owned by the public, and when we speak of the public realm, we are speaking in large measure of streets. If we can develop and design streets so that they are wonderful, fulfilling places to be, community-building places, attractive public places for all people of cities and neighborhoods, then we will have successfully designed about one-third of the city directly and will have had an immense impact on the rest."*

The sidewalk is, as she called them, "the main public places of the city" and "its most vital organ." In a similar vein, Jan Gehl (1987) calls attention to the *life between buildings.* It is in the spaces between architecture where the social connections of inhabitants are created and reaffirmed and where the space of movement coexists with the city's social life. These aspects form inter-related patterns of movement and activity to which buildings, in turn, respond. To Kevin Lynch (1960), this *life* also imbues these spaces with meaning, whereby they become "imageable," containing spatial qualities that strengthen the attention and meaning of urban space to its inhabitants. Such spaces establish memorable relationships among buildings, paths, features, and amenities due to their proximity to one another or the geometric configuration of the space itself.

Ultimately, these concepts are intended to influence the practice of *place*-making or the shaping of urban spaces that support the public activities mentioned. Designers are often unaware of their own biases and roots as influencing their decision-making (A. Jacobs & Appleyard, 1987). The profession of urban planning would be helped by a renewed quest for the elements of "good city form," where the planning of the physical city is concentrated on *people* and *place* (Emily Talen & Ellis, 2015). For designers and planners, normative limits such as time and budgets are often drivers of the opportunities available to improve urban space. However, they are often exacerbated by a lack of understanding of how both users and non-users perceive them and the characteristics of users (Chidster, 1989). Such an endeavor would have to deal with the complexities of aesthetic, ethical, and political theory to secure its foundations and, as such, require domain-specific ontologies unique to the social and physical milieu of urbanism.

## 3. IMAGERY AS URBAN DATA

Imagery had enormous agency in being evidence of human activity in public space. Our contemporary era has seen the ubiquity of cameras as tools used to documentation of social ills and injustice. Videos of vegetable seller Mohamed Bouazizi setting himself on fire in protest in Tunisia (Noueihed, 2011) or of George Floyd gasping "I can't breathe" as police offers aggressively restrained him (J. Collins, 2020) have served as evidence for protests demanding justice and change. Yet, the use of photographic imagery as a data source for learning and exploring urban space dates as far back as the technology itself.

Louis Daguerre's *View of the Boulevard du Temple*, created in 1938, was the earliest photograph to include people in a busy street. Although it depicts two men on a street corner, they appear only because one has his shoes polished by the other and thus required them to be stationary. The long exposure required left no trace of moving traffic and other people. In the 1860s, the nascent art of urban photography became a political tool for social and urban reform. In Paris, Charles Meville's commissions to photo-document the modernization efforts of Baron Georges-Eugène Haussmann recorded both dirty and cramped "before" and grand, triumphalist "after" vignettes of the city. Melville's photographs have been the fodder for debates on whether they served more than an archive of a foregone Paris on behalf of the city by also acting as propagandist justifications for the razing of many neighborhoods (Dahlberg, 2015). Historian Shelley Rice's (1997) critiques of Meville's repetitious and "clinical" streetscapes point to the photographer's "passivity" toward the grand transformations of the city. On the other side of the Atlantic, Jacob Riis, while working as a police reporter for the New York Tribune, documented the slum conditions on the Lower East Side of Manhattan. Using lantern slide shows, illustrated articles, and printed books, Riis harnessed the power of photographic imagery as evidence of the need for housing and social reform, which galvanized the politically connected gentile and policymakers to action and reform.

Where once rare, the twentieth century saw the increased ubiquity of cameras which served in the modernist era as tools that could provide an unblinking empiricism in the documentation of human behavior. Powered by the increased availability to take photos from aircraft and satellites, planners and architects were drawn to the newly available planar perspective afforded by flight as a way to document the formal development of the city. For Le Corbusier (1935), the eye of the airplane offered a "pitiless", objective record of reality, and for many planners in the United States and Europe in the post-war era, aerial photographs served as evidence of urban blight. This documentation provided the opportunity of seeing the world anew, but also as confirmation for the need to *replan* (Hinchcliffe, 2010). In their seminal work, *Learning from Las Vegas*, Venturi and Scott Brown (1972) experimented with various visual media to explore ways of representing American urban form in the late-1960s. As a critical exploration about the automotive orientation and iconographic expressiveness of Las Vegas, their Yale research group explored the city using photography and film as both mechanisms for data collection and representation. Notably, their post-modernist approach adopted "drive-by photography" taken from inside moving cars, a technique borrowed from the artist Ed Ruscha, as a means of contextual representation.

Taken together, the imagery serving as data also produces meaning as understood by people. In investigating urbanism, what is learned about cities and their inhabitants is both mediated and shaped by the characteristics and limitations of the media and the mode of production (Azar et al., 2021). While historic precedent relies on the curation of the framing and representation, emerging technologies are increasingly less reliant on humans—images are increasingly being made by machines, for machines (Paglan, 2016). This computational process forms a shift in how the production of urban knowledge, vis-à-vis enumeration, data, or models, is being generated and understood, while remaining potentially invisible to the human altogether.

## 4. IMAGE-BASED ANALYSIS IN THE "SMART CITY"

From the horse to the automobile to the telephone, the physical city, as a social artifact, has long evolved in response to prevalent technologies of the day. Today, the rise of the so-called "smart city" introduces two parallel technological interventions: a digitally connected citizenry possessing connected devices such as mobile phones and the ubiquitous instrumentation of the physical artifice of the city through the "internet of things." Pervasive sensing enabled by both has created the ability to dynamically sense, analyze and understand individual mobility traces more quickly and accumulate detailed knowledge over time to see patterns and trends.

This "big data" approach—having access to large volume datasets to study phenomena and their dynamics—augments the process by which urban space is designed, developed and evaluated, and offers opportunities for data-driven analysis and design of the built environment. McLuhan foresaw technologies serving as civic thermostats "to pattern life in ways that will optimize human awareness" (Norden, 1969). He said, "already, it is technologically feasible to employ the computer to program societies in beneficial ways." He stressed that "the programming of societies could actually be conducted quite constructively and humanistically." Greenfield (2013) comments that "the final intent of all this... is to make every unfolding process of the city visible […], to render the previously opaque or indeterminate not merely knowable but actionable."

Despite its complexity, the desire to make understandable the complex nature of vibrant cities has been both a dream of Euclidean modernists and those who pursue an organic method of urban planning. For instance, for urbanist Jane Jacobs, "the variables are many, but they are not helter-skelter; they are interrelated into an organic whole." Jacobs was talking about Hudson Street in 1960s-era Manhattan (J. Jacobs, 1970), but the same could be said of any street in any city today. The critical difference is that today, digital data allows us to describe and analyze Jacobs's "interrelated variables" better than she could ever have imagined. Big Data may lead to the discovery of fascinating dependencies and intimate knowledge. (Jemielniak, 2020)

As city officials, urbanists, and urban planners analyze the figurative reams of data available; imagery has also been leveraged as a means of understanding larger urban dynamics. For instance, Girardin et al. (2008) used individually-uploaded photographs on the online photograph sharing platform Flickr to assess the different visitation footprints in Barcelona of domestic, Spanish tourists, and international Britons visiting the country. To do so, the researchers analyzed the underlying exchangeable image file (EXIF) that stores metadata created by the camera and user-volunteered information such as locations, profile demographics, including the countries and cities where the photographer was from, tags and descriptions. The fusion of this information provided insight into how different groups may use a city differently (Offenhuber et al., 2013): the British tend to stay to primary, famed tourist sites such as La Ramba, while Spaniards were willing to go farther afield. It also reaffirmed notions of the "imageable city,"[1] exposing locations that attracted photographers' attention (monuments, churches, public spaces, for instance) were exposed, and where the absence of images may reveal locales considered more *introverted*.

Although the use of volunteered imagery is still commonly used, Street View images have provided a dataset that is both immensely comprehensive in its coverage (thanks to the deep financial resources of its creator company, Google) and consistent in its format. This undertaking was driven by co-founder Larry Page's belief that this type of street-level imagery contained a tremendous amount of information that could be organized, made available, and mined (Anguelov et al., 2010). By 2017, the company captured 16 million kilometers of Street View imagery across 83 countries (Ackerman, 2017). This powerful dataset, albeit through restrictive commercial and proprietary means, has allowed researchers to investigate street-level dynamics and changes of urban space at a massive, aggregated scale from understanding the prevalence of tree canopies and green space (Li et al., 2015) to pedestrian counts (Yin et al., 2015).

---

[1] "Imageable" here is taken to mean a cognitive, memory-based image as was used in the previous reference of Lynch's work.

However, several projects have sought to use this dataset to assess more ephemeral or culturally specific attributes about these spaces. Salesses et al. (2013) used the street-level images in an online survey of people's reactions to whether a space looked more or less safe than another to index the perceived safety of a specific location in many cities around the world, based on visible characteristics such as street width, program use, apparent blight, among many other formal attributes seen in the photograph. Taken together, Naik & Philipoom (2014) use machine learning algorithms to scale the survey results into an index applicable to many other cities around the world. The EthinCITY Linguistic Landscape project from the Spatial Analysis Lab (2019) captured 9 million textual signs on the streets of Los Angeles—ranging from official street names to business signs and advertisements. By extracting the text and language of the message and geocoding the words and languages, the team assessed the languages used to the parcel level. What emerged is a dynamic map that reflects the cultural diversity of tableau that traditional government sources of data cannot collect, including small, ethnic establishments. In addition to macro-scale findings such as there being not just one "Chinatown" in the city but five smaller ones, they could redefine notions of ethnic enclaves as static neighborhoods and instead explore the dynamic, overlapping, and interspersed nature of urban cultural diversity.

While user-generated imagery comes at extraordinary velocity, volume, and veracity, a general lack of standardization of what is captured[2] and when limits the potential for a deeper inquiry into specific locales. For instance, while there are over 2 million photographs of "Disneyland" in California on Flickr, there is only one of "Captain Kidd's Family Dining" located just across the street from the entrance to the theme park. Moreover, while Street View images remedy the aspects of standardization at a massive scale, the enormous cost of capturing the images limits the potential to understand the whole dynamic nature of cities in real-time. However, the emergence of smart city technologies through placed sensors provides opportunities for deeper urban inquiry that addresses the standardization and volume concerns while providing the opportunity to understand places in real-time and over longer durations.

**4.1. Automated Eyes on the Street**

With the emergence of the smart cities movement, much investment in sensing has focused on the domain of computer vision, advanced machine learning (or deep learning), and more specifically, Convolutional Neural Networks (CNN) to perform analytics on a single image or a series of images. This artificial intelligence is concerned with the automatic analysis and extraction of useful information from these images or videos. In other domains, computer vision has opened new frontiers in disease treatment and diagnosis, agricultural and industrial efficiency, robotic navigation and self-driving vehicles, and more. In cities, this automated technology is being used in wildlife tracking, intelligent traffic management, pedestrian flow monitoring, public safety, and infrastructure management (Kwet, 2020).

This boom is following a concomitant growth in the number of surveillance cameras being placed in the built environment for various reasons and by a multitude of public and private parties (Lin & Purnell, 2019).
It is estimated that the number of surveillance cameras globally will exceed one billion by the end of 2021, with over 85 million in the United States alone, rivaling China's per person camera penetration rate (IHS Markit, 2019). While cameras on their own lack the overall ability to do the computational performance being described, it is estimated that 350 million of these situated cameras will possess built-in technology to allow for artificial intelligence in 2025, with more than 65 per cent of cameras shipped in that year will come with at least one AI chipset (ABI Research, 2021).

---

[2] These together form the commonly used "four V's" of big data: velocity, veracity, volume and variety.

To enumerate and describe the behaviors of inhabitants, these smart devices apply a domain of vision-based machine learning called "action recognition," which seeks to extract peoples' identifiable physical features, acts, and behaviors. Generally, all computer vision approaches generate data abstractions from the captured images and videos, thereby reducing the image to a series of numerical matrices that can be more easily analyzed.[3] These matrices are produced by assessing patterns in the red-green-blue (RBG) pixel values of a digital image (and depth information, if available). They are compared against pre-trained models developed from the analysis of prior imagery. The creation of this pre-trained model is vital as it provides the foundation for comparison, much like how humans learn how to identify objects: we learn by establishing priors—formal characteristic, visual appearance, defining features, or other attributes—to be able to hypothesize about the nature of something new. To get to the point of accurate enumeration, enormous data is required to train the statistical and deep learning models to ascertain human action from noise.

These processes are often ambivalent to the identification of a human as a specific individual unless they are specifically focused on that task, such as through the use of facial recognition technology. Generally, the first step in identifying human action is to identify features in an image that statistically resembles the figure of a person. Here, one can think of this abstracted figure as a blob, box, or skeleton (Figure 1). Each of these offers different tradeoffs in the computational resources and statistical strength but represents humans in different ways.

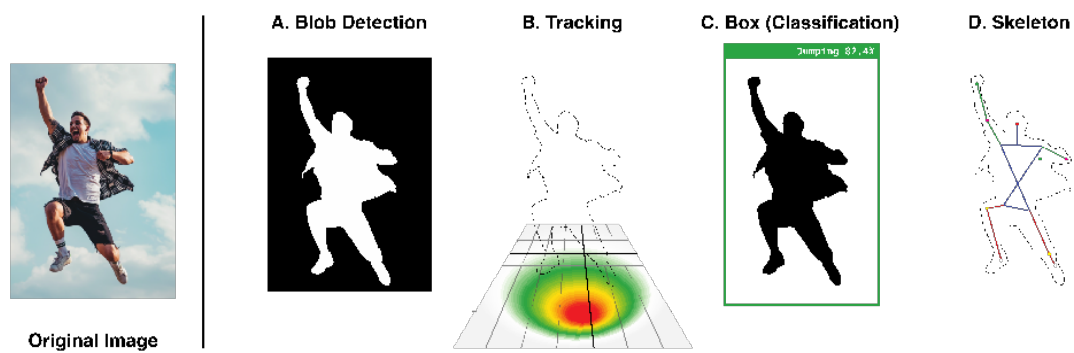

*Figure 1. How a computer interprets subjects and actions in an image. (Original photograph credit: Chetan, 2019.)*

The computationally simplest method is blob detection, whereby an algorithm can assess a figure in the foreground of a video as being distinct from the background by comparing attributes that change between frames. This approach toward blob detection is very frequently used in describing movement flows. In Figure 1, we simply identify the hotspot of where the person is or has been as a measure of where the figure has been relative to the camera's view, establishing the background frame. The dichotomous nature of this foreground-background comparison can be used to make the assignment of actions computationally more efficient by focusing the more intense box or skeleton analysis on a smaller portion of the total image, however. Within this subset, an algorithm can do one of several things: compare track this *boxed subset* within the frame, compare the RBG pixel values against the trained models (classification), or further reduce the figure to the relationships of certain parts as probable limbs or critical physical features vis-à-vis a digital, topological skeleton. Ultimately, it is the comparison of these two reductions, the boxed pixel values or the skeleton, that is used to assess human action algorithmically. These actions can be categorized according to the complexity of activity: primitive, single person, interaction, and group (Al-Faris et al., 2020), and form the foundation for the computer assessment of activity. The combination of these methods and categories allows us to identify individuals' activities in urban space.

---

[3] While this paper will not comprehensively review the technology and its application depth, other papers have sought to categorize various approaches. See *Ibrahim et al., 2020*, for instance.

## 4.2. The Players and Builders

The use of the term smart cities has been characterized as incorporating computational tools into the operations and management of municipalities and regions. However, this also brought a different paradigm of urban improvement as the private sector played a more significant role in city-building. Many cities saw the implementation of these technologies as two-fold endeavors: a means of economic development in an increasingly globally competitive marketplace (Zukin, 2020), and in light of diminished resources, a means of outsourced services management. Particularly in light of the inability of city bureaucrats to develop or implement new data collection means, greater attention has been paid to technology companies, startups, and academics to create and implement these new technologies within the urban domain.

Fundamental to planning or policymaking is the restructuring of complexity to a discrete set of standardized measurements that are understandable by those shaping the environment. Normatively, these measures are drawn from a desire to benchmark the use or inhabitation of public space. Much of the development of these technologies have operated at two scales: the infrastructural, driven by transportation and mobility efficiency, and the architectural, driven to quantify the economic productivity of a space.

Although primarily focused on the level of service or throughput of a street, much attention has been paid toward the management of road infrastructure. With increased demand on existing and aging infrastructure leading to congestion and economic tolls, many entities see opportunities to develop technology to "solve" traffic. Many universities (Yang et al., 2020), startups such as Miovision, CurrexVision, and vivacitylabs, and incumbent automotive companies such as Honda have developed different algorithms and sensor packages to count and track automotive traffic on roadways, each with slightly distinct approaches informed by the countries from which the companies are from. Automotive companies are also seeking to leverage the data aggregated from individual vehicle's sensor suites (Massaro et al., 2017) including the onboard, external cameras to analyze for future products and to use as derivative data that can be monetized.

Similar technologies have also been used to analyze interior spaces, particularly in retail or commercial environments. Companies like Indoor Atlas and Brickstream are applying similar approaches as with car traffic, including understanding common paths and behaviors to understand consumer patterns in discrete spaces. RetailNext, in addition to algorithmic counting tracking of occupants and heatmap generation, fuses image data with cellphone tracking to know how these patterns differentiate between men, women, and children and track repeat visitors to a space.

The public adoption of online shopping has required better management of these areas of the street and so valuable is this piece of real estate that cities are investing in systems to increase access to this space. The World Economic Forum (2020) estimates that the number of commercial delivery vehicles will increase by 36% in inner cities, globally. Using similar technologies as the transportation measurement companies, companies like Automotus and curbFlow focus on the management and monetization of the curb to increase the curb's efficient use for deliveries, parking, and repair.

In public, human-oriented spaces such as sidewalks and plazas, companies like Numina and Placemeter have sought to enumerate individuals on foot or bicycle. Driven by concerns about privacy, Natix and Numina has taken a different approach than many other companies to intentionally reduce the amount of data it assesses and does it within the sensor through an edge computing framework, with the intent to put privacy into precedence. Placemeter, now part of Netgear's Arlo home monitoring brand, used off-the-shelf, consumer-grade cameras to quantify and categorize people, bicycles, motorcycles and vehicles. With the City of Paris, Placemeter piloted projects on the enumeration of a public plaza and pools. While much development into the technology has been done into issues of crowding dynamics (usually within the rhetoric of public safety) and level of service, little work has been done to precisely understand the complexity of pedestrian behavior. While research is emerging into understanding non-commuting behavior (Seer et al., 2014; Sun et al., 2020)—that is to say, ambulating or non-intentional travel—technologies for understanding social behaviors are still nascent.

With the creation of the LinkNYC network of upwards of 7,500 digital kiosks built by Intersection (a subsidiary of Sidewalk Labs, which is a sister company to Google, and a subsidiary of Alphabet), many residents were concerned with the potential invasion of privacy through citywide, digital tracking, and issues with ubiquitous security surveillance that would result (Kofman, 2018). Originally proposed as a twenty-first-century replacement to the ubiquitous payphone, the new kiosks would provide free Wi-Fi and digital signage and included a suite of three cameras and thirty sensors and heightened sight lines above the crowds and onto the streets below. While these technologies were touted as opportunities to move into infrastructure management and pedestrian counting, concerns arose around the program's vast and indefinite data retention and the possibilities for unwarranted NYPD surveillance (ACLU, 2016).

While not fundamentally opposed to the potential to understand public behavior, these camera technologies can also be used for policing and security purposes at large using the same algorithms, whose ethical rationales have conflicted with civil libertarians. This issue arises when the desire to scale and monetize every facet of the artificial intelligence and sensing platform manifests itself into multiple products; when services used for the public good can be easily adapted for surveillance practices. For instance, cities increasingly turn to facial recognition to identify individuals with outstanding warrants or pose specific threats. The Singaporean police force, for example, has installed nearly 80,000 cameras in key public areas, with each camera having the capability to run real-time video analytics to detect potential criminal threats (Lin & Purnell, 2019). Large companies like Verizon, Amazon, IBM, and Microsoft have readily available, scalable technologies that offer police and security agencies the ability to capture and fuse various image sources to provide real-time situational awareness and identification of individuals in a crowd. The New York Police Department, for example, partnered with Microsoft to equip cameras with the technology to, they claim, identify crime based on "potentially suspicious body language (Stanley, 2019)."

However, in the contentious public debate about using these technologies, the concerns span public and privately-owned digital infrastructure. In addition to public networks, cities are also developing so-called "plug-in" platforms where business owners can connect their privately-owned, internet-connected cameras to city-operated camera networks. Most notable of these is Project Greenlight in Detroit, although similar networks exist across the United States. Their automation systems are built upon facial recognition technology from biometrics system company DataWorks Plus. The project uses the company's Face Plus video surveillance product. It compares captured faces to a database of over 500,000 mugshots with additional access to a statewide database that includes drivers' license photographs (Garvie & Moy, 2019). In 2021, the network has grown to over 700 cameras, but cities like Chicago and New York have tens of thousands of networked surveillance devices (Kwet, 2020). Despite concerns from individual homeowners, Amazon has also made data from its branded doorbell cameras to 400 police forces (Harwell, 2019).

The use of these technologies has come with much criticism, including from the industry itself (Shepardson, 2020), for the lack of transparency around the use of these technologies—including their questionable accuracy (Hill, 2020; Lee, 2020)—and the ethical concerns around their use. Using algorithms to label people based on race or ethnicity has become relatively easy from a technology standpoint. Both IBM and Microsoft readily advertise their services to sort people into broad groups, including by race (*Attribute Detection with Body Camera Analytics*, 2020). With reports that this technology has been used for tracking of to track and control Uighurs individuals in China (Mozer, 2019) and Black Lives Matter protesters in the United States (Selinger & Fox Cahn, 2020), many are questioning the appropriateness of these technologies in cities.

The same technologies used to identify faces are also being used to identify features of the face for advertising and retail purposes. Emerging technologies read facial characteristics to measure different types of engagement with media, such as measuring attention (Picard, 1995). Startups like RealEyes, SightCorp, and Kairos offer technologies to pinpoint the coordinates and directionality of a person's gaze, relating it to capturing and holding an individual's attention. Going further, Affectiva is tuning to the characteristics of the face as it relates to affect, measuring one's emotional state. Their algorithms have been tuned through the analysis of 7.5 million faces from 87 countries, mostly collected from opt-in recordings of people watching TV or driving their daily commute. While their applications are clear concerning enumerating commercial activity, these datasets can also be tuned toward understanding facets of the built environment at large, including the emotional landscape of cities. For instance, an application of affective computing applied to communities explored the overall emotional happiness around the university campus of MIT by reading the faces of passers-by and finding the variances by department, building, and time of semester (Hernandez et al., 2012).

**4.3. The Underlying Data and the Underlying Problem**

The generation of identities, tracks, or outputs is incumbent on creating models that can interpret the live stream of real-world data. Therefore, a dataset is required to both train and validate these models and must include additional annotations and metadata to ascertain meaning from the results. Due to the emerging nature of these technologies, only a handful of datasets account for the vast variety of model-creation. However, a growing number of specialized datasets are also becoming available to the research and development community.

Among the most commonly used training dataset for computer vision identification are Imagenet (Deng et al., 2009) which includes approximately 14 million hand-annotated images and 20,000 categories, and Open Images, with its nine million images and nearly 20,000 human-verified classes (Krasin et al., 2017). At the same time, many datasets have been created specifically to train models on human action (Ofli et al., 2013), most derived opportunistically from existing sources such as film and television or online video-sharing platforms (Idrees et al., 2017; Patron-Perez et al., 2012; Smaira et al., 2020). The Kinetics-700 dataset, for instance, used 650,000 YouTube video clips to generate a dataset covering 700 human actions. Business interests are also driving the creation of specific datasets to improve the abilities of computer vision in the marketplace. The Affectiva-MIT Facial Expression Detection dataset (McDuff, El Kaliouby, Senechal, et al., 2013) is a provocative image collection of facial emotions of people from around the world, enabling researchers and companies to measure affect, with applications towards retail, media and advertising (McDuff, El Kaliouby, Demirdjian, et al., 2013). The increased interest in self-driving vehicles led to creating the Cityscapes dataset (Cordts et al., 2016), which annotated 25,000 images specific to urban cityscapes taken from the street.

While the latter shows promise to shape a digital understanding of urban spaces, the focus on vehicles narrowly focuses the image set on the spatial perspective of the driver and a car's navigation through this singular aspect of the city. The Cityscape does identify people and 30 classes ranging from "sky" to recognizing different types of street objects and vehicles, but only as it informs a vehicles' ability to navigate a street and not a proper understanding of how an urban space operates. Additionally, being generated from the imagery of only German cities, it may lack a diversity of potential, infrastructural histories and standards, and mobility options.

This point poses a critical question within the space of urbanism: can they account for cities' real experiential and social diversity? Of course, one cannot assume that the compilation of data with varying levels of detail and depth will always be problem-free (Brannen, 2005). The challenge in creating these datasets is the embedded and implicit values and assumptions made by those who both organize and annotate the images.

Firstly, the orientation towards universal applicability of these datasets creates an ontology that privileges generalities over specificity where no agreed upon standard may exist. Societal appreciation of architectural styles or the role it plays in broad society, for instance, varies greatly between cultures and context (Berlyn, 1971; Kubo et al., 2010).

Secondly, these datasets built from the opportunistic collection of imagery, such as ImageNet and Kinetics-700, may omit specific populations' experiences because of more significant societal inequities. The gender and racial inequities (R. L. Collins, 2011; Desmond & Danilewicz, 2010) and a lack of geographic diversity (Shankar et al., 2017) in television and media may be reinforced by the derivative use these datasets in the training of new algorithms unless ameliorated. For instance, the largest share of users and content on YouTube originate from the United States. Similarly, the United States exerts enormous global soft power in the production of television and media. Any algorithm trained on that dataset will perform worse on non-American contexts with increasing error as a culture is more distinct from that standard. For digital images to be useful, they must be organized in accordance with some type of knowledge system (Pasquinelli, 2015). As this data disappears into the black box of the algorithm and manifests itself as the building block for policy and the physical reshaping of our cities, it is incumbent to question whether these existing datasets and paradigms adequately describe urbanism.

## 5. MEANING IN SPACE

Innovations in computing avail new avenues for urban research, but critical considerations remain regarding the use of these technologies within an urbanism discourse. As Duarte and DeSouza (2020) argue, urban technologies must consider more fundamentally how the epistemologies behind extensive data methodologies as well as how they shape ontologies and heuristics about cities, as ultimately there will be lasting transformations to both society and space due to the specific responsibility of urban planning in shaping communities.

Among the epistemological standards with computer vision technologies is the orientation towards reductivism or simplified numeric representations (Dreyfus, 1992). In itself, this approach has vast implications for how residents and policymakers understand the built environment (Winner, 2017). While this reductivism can produce a generalized intelligence about a location that may lead to prediction, this reductivism is often uninterested in the ephemeral qualities that make cities unique. Companies that are developing these technologies strive for scalability or the universal applicability of their products. This efficiency strategy is opposed to considering unique or distinguishing factors unique to a place, including the inherent complexity of such environments (Gershenson, 2013; Gill, 2020). It thus prioritizes simple-to-measure factors such as counts or duration over nuanced observation of sociability or idiosyncratic behavior. As Czarniawska (1992) frames this focus, the current paradigm of urban observation excludes the "anthropological frame of mind" that more seeks to explain these behaviors.

This conflict of definition parallel Clifford Geertz's (1973) arguments for an interpretive approach to understanding culture. To Geertz, the existing paradigm of anthropological research was through *thin descriptions* derived from which includes surface-level observations of behavior. Similar to the superficial enumeration of many urban technologies, these observations lack the context and interpretive turn that define *thick descriptions.* He analogizes the difference through identification of a wink between two children, the thin description. However, the meaning behind the wink—be it coded communication between them, a random biological twitch, or an attempt at seeking attention—can only be understood through think descriptions. For urbanism, a similar a framework could be understood in everyday activities. For instance, someone holding out their hand in a park may be a gesture to a friend while the same gesture in the street may be the hailing of a taxi.

The creation of the metrics by which cities are documented can also distort toward one type of description. There is also a worry that the ease of thin description through both data collection and analytics runs a risk of privileging infrastructural efficiency over the more complex social promotion. The apparent focus in the mobility domain toward level of service also bias interpretations of cities toward thin descriptions, as they premise cities on the sole metric of infrastructural efficiency. The relative ease of this type of data collection has implications on how the cities are described, with the risk that they are seen as optimization challenges versus social environments.

This conflict is fundamental to the emerging practice of big data research, let alone when embedded within complex social environments such as cities. Geertz's position with the derivation of thick descriptions sought to evolve ethnographic research toward the ascertainment of more significant meaning. Nevertheless, it relied on the existing models of practice as a foundation for criticism. However, in urban science research, like the computational social sciences on which it is founded, the canons of practice are still being formed (Jemielniak, 2020; Rossman & Rallis, 2017). Thus, the critical observation of urban phenomena is caught uncomfortably unformed between the data science discourse—where the prevailing axiom accepts the unknowability of models and analytics performed within the analogous black box are worth the gains in accuracy from such processes—and that of the interpretive anthropological work.

Lessons in performing observations in this liminal space between reductivism and interpretation may be found from pre-computer research. In the pursuit of documenting public space as part of the *Street Life Project*, William H. Whyte (1980) notably used Super 8 film to understand the use of plazas and public spaces. Whyte and his associates recorded and analyzed hundreds of hours of time-lapse film to assess variation and regularity in the behavior of anonymous pedestrians in just a handful of cases. His research was performed to provide data critical to measuring activity patterns in urban spaces, which did not exist at the time. Whyte's investigations are among the earliest attempts at using new technologies to surveil and understand human activities at scale. While the process was still incredibly manual, as researchers needed to comb hours of video manually, it was an attempt to document an entire public square as a whole and an example of a technology that was not yet readily available to the public. In doing so, the project revealed both generalized behaviors, such as the conclusion that park use was directly related to the amount of "sittable space" and not shape or size as previously thought, and idiosyncratic behaviors of individuals, such as gender differences in seat and location choice.

The dual nature of his research on ascertaining thin, generalized patterns of behavior and thicker, contextually derived considerations allowed Whyte to discover practices that would inform zoning regulations and propose interventions. For instance, a key characteristic identified for plazas that succeeded as popular gathering spots was the presence of a variety and abundance of places to sit, including benches, movable chairs, ledges, and steps. They also found that physical characteristics, including tree canopies, water features, public art, and food vendors, all played a role in attracting people to urban plazas and parks. These attributes played a role in creating positive feedback loops for bringing people together, while streets observed with blank walls and devoid of shops, windows, or doors saw little activity. In Whyte's words, "what attracts people most, it would appear, is other people." While these findings are taken for granted in the twenty-first century, Whyte's findings broke with contemporary paradigms of modernist urban design. These findings would notably lead to New York City passing zoning regulations that required plazas and publicly-owned public spaces to no more than three feet above or three feet below street level to allow for visibility and easy access, and the complete revitalization of Bryant Park, which in the 1980s saw little activity, to increase its visibility and to add more seating, including its now-familiar moveable furniture. Today, the park is popular year-round with more than 1,000 movable chairs plus several cafe kiosks and many scheduled events.

### 5.1. The Conflicts in Meaning

There was much optimism for new opportunities for digitally-mediated and more open civic governance (Goldsmith & Crawford, 2014). However, the mismatched objectives of neo-liberal development and the scale-driven focus of industry and the slower, discursive, community-oriented democratic process of urban management in many cases yielded incompatible metrics and conflicting outcomes. While it is not in dispute that the ability to understand and enumerate how residents used urban spaces could lead to more appropriately responsive interventions, the more considerable utility of these datasets to urbanists is questioned. Fundamental to this conflict is how each party saw the challenge of enumeration and what would entail defining the "use" of public space.

Similarly, tradeoffs with computational intensity limit the fidelity by which technologists can process the imagery. As imagery processing requires immense computing power, and many companies rely on weaker edge computing platforms to preserve privacy, the scale, detail, and precision of what can be enumerated is directly related to the number of computational resources available. These limitations are present at all points in the technology: from the captured image's pixel resolution, to the storage and transmission of the image, to the amount of processing power available to convert the image into a useable measurement. As such, many purported benefits of particular technologies are mitigated by such computation's cost and physical demands.

Ultimately a series of tradeoffs, the interrelationship between capacity limitations, societal impacts, and the thickness of its meaning offers a way to evaluate how citizens can evaluate the appropriateness of computer vision or any smart cities technology. In Figure 2, we evaluate the companies previously mentioned to organize the tradeoffs and limitations on each of these three metrics. Taken together, they offer a way for citizens to both assess the adequacy of various technologies for their community and find ontological clusters of technologies that share similar opportunities and constraints across domains.

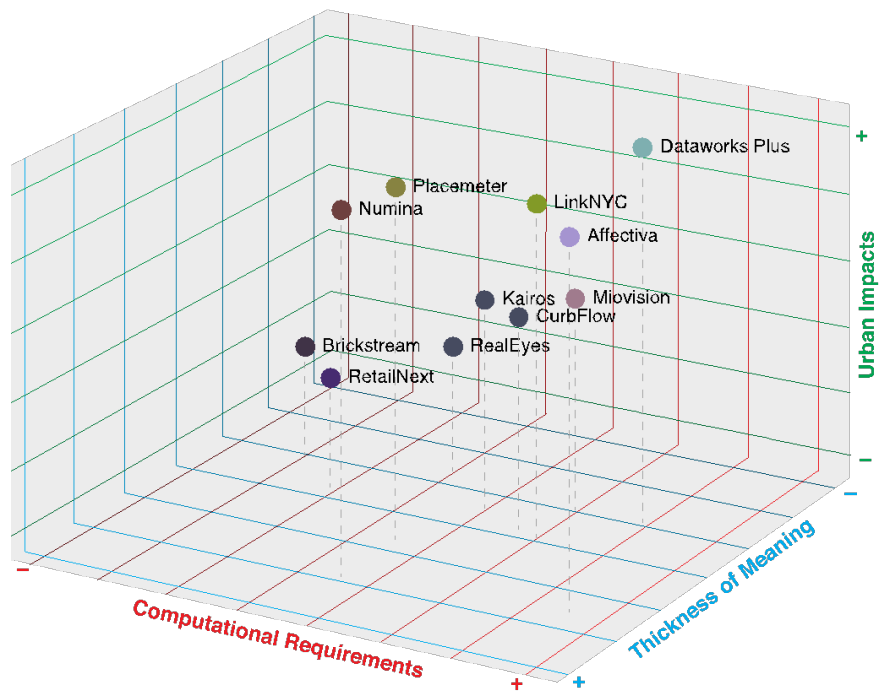

*Figure 2. Evaluating the meaning, impacts and computational costs of various technologies.*

This taxonomy of tradeoffs also offers a lens by which citizens can evaluate the potential societal impacts of computer vision technologies. Where certain activities such as the maintenance of a sidewalk bear few stakes for society at large, high-fidelity and detail are likely unnecessary, and costs associated with processing high resolution, real-time images are likely unnecessary. However, these technologies' use in policing has high implications for the citizens when questions of justice are involved and requires extraordinary precision and accuracy. For instance, the high false positive rate of the Detroit facial recognition program, where the police chief points to a 96% error rate and reflects the inadequacies of the technologies and algorithms (Lee, 2020), should give pause in how readily citizens should accept the meaning of its outputs.

## 6. REFRAMING THE METHODS TOWARD URBAN-SEMANTICS

Many of the vision-based smart cities technologies that seek to understand how people use space lack specificity for how people use urban space within urbanism discourse. This presents an opportunity for computer vision research to develop semantically specific approaches that are appropriate and contextually informed to the nuances of urban studies. However, what would an urban-sematic computer vision look like? To relate to Geertz's writing, at present, the state of computer vision is proficient in identifying *winks* but does very poorly in identifying what those discrete objects may mean within a city.

The state of the art of technology allows for a computer to identify objects and actions in isolation. However, when accurate, the definitions are thin in their descriptions without a situation to contextualize the assessment.
In Figure 3, four photographs of "people sitting on a bench" were run through the ImSitu object recognition algorithm (Yatskar et al., 2016) to identify the images through the algorithm's perspective. The algorithm attempts situation recognition, by which the algorithm provides a concise summary by including the main activity, the actors, and roles. The algorithm was trained on FrameNet, which contained over 125,000 images and 200,000 unique situations. The same images were identified by individuals working in urban design and planning, and the most frequent description was used in the image, although the human descriptions varied little.

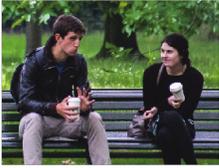

*Figure 3. Results of ImSitu algorithm versus human identification of four images of people sitting on a bench. (Photograph credits: hjl, 2012; byronv2, 2019, 2020a, 2020b)*

The findings showed fairly accurate thin descriptions across the four images, although the confidence varied due to the image quality and color differences on each image. While it was interesting that the algorithm could identify individuals sitting on a bench, the nuance of what each group was doing was missed. This could be due to a variety of reasons, including the lack of lexical data on those specific activities, the quality or perspective of the image, or improper training of the dataset. In any case, the algorithm could not achieve the definition of the already-reductive human descriptions and demonstrates the current challenge of assessing thicker descriptions of urban activities.

Recognizing the limitations of state of the art in computer vision, we highlight potential methodologies by which technology developers can gain a thicker understanding of activities in urban spaces. The challenge with using algorithmic black boxes is that the processes from which outcomes are computed are unknown. However, understanding the general mechanics of how algorithms operate may open opportunities to contextualize computer vision toward urban-specific contexts.

Firstly, when we can consider how models are derived and the generalized libraries commonly used for training. We can derive urban-specific identification through *a priori* contextualization by creating a spatially specific corpus of images containing descriptions of and action annotations calibrated to urbanism. That is, an unambiguously urban dataset must be arranged in a taxonomic form so that an algorithm might deem what is and is not relevant information specific to a particular context. A disadvantage of this approach is the necessity to plan in advance the classes and categories of urban-semantic annotations or relying on a large group of human annotators trained toward specific city-focused labels of footage.

An alternative is to consider an *a posteriori* approach. Here, we leverage the strengths of computer vision algorithms to find common patterns from imagery, and an algorithm clusters similar activity. Humans could then verify and label tag these clusters with thicker descriptions after the clusters were found. While technically feasible, unsupervised action identification from the footage is new (O'Hara et al., 2011; Soomro & Shah, 2017) but can allow urbanists to revisit the world of individuals like Whyte to the computer could find that humans could not.

Within a societal context of cities, there are also questions about these technologies' relationship with the public. Residents are not strictly customers nor users of these services and may not have any option to opt-out of these technologies. Despite the risk of these black-box processes reinforcing the detrimental *status quo*, or worse, furthering pre-existing bias, how these technologies are created and therefore drive planning, policy, and design decisions often lack the input of those whom they will impact. The Boston Beta Blocks program was created as a policy-based mechanism to pilot technologies and create a platform for community engagement and empowerment. In addition to civic experimentation with technologies, the program also organizes educational workshops and events with the platform providers, whether or not they are considered for procurement (Mayor's Office for New Urban Mechanics, 2018).

Under the civic experimentation mandate, new technologies are piloted in the open in pre-selected areas where the community has mechanisms for feedback, offering transparency and citizen oversight into selecting, testing, and creating success metrics. The residents are invited to co-generate with the city and the companies the values around civic and privacy concerns. They also provide oversight into the processes and policies that govern these experiments. As a result, the social milieu around the implementation of technologies is contextualized to the people, needs, place, and time of that specific community.

## 7. CONCLUSION

The current orientation of computer vision technologies has been inwardly focused on its own development and toward broad generalizability of its applications. However, as the adage goes, "when something is good at everything, it is the best at nothing." When these technologies are implemented within the complex milieu of cities, the application thus far has erred toward reductive quantifications at the sacrifice of the dynamic characteristics of public space that draw billions of people to live, work, and play. Precisely because of these dynamic characteristics, the conception and development of these tools should be reconsidered to appreciate the idiosyncrasies of these spaces.

Drawing from precedent conversations in anthropology and the social sciences, urban technologists must move algorithmic enumeration away from simplistic ontologies toward thick descriptions that better capture the dynamism of cities. Like Whyte, the interrelationship between enumeration, description, and interpretation is vital to drawing conclusions about the urban spaces. As the proliferation of these technologies persists, mechanisms to consider bias in the recording, training, development and use of these datasets and algorithms are vital, especially when the benefits may be inequitably born by inhabitants, and because the role these analytics may play in the shaping of the built environment and its policies.

While imperfect, these proposed approaches allow urbanist to move slightly away from what David Hand (2020) considers "dark data," the data that is inaccessible from current tools and do not fit within existing methods, but still can influence the decisions and policies that may result. These approaches allow urbanists to "hard work of theory" to critically examine the ontological and epistemological frameworks that exist with the use of these technologies, and reorient the practice toward metrics that relate to the social life of cities and away from reductive, service-oriented quantifications (Pickles, 1997).